\documentclass[12pt,preprint]{aastex}

\def\teff{${\rm T}_{\rm eff}$}
\def\vturb{${\rm v}_{\rm turb}$}

\usepackage[latin9]{inputenc}

\begin{document}

\title{A chemical {\sl trompe-l'\oe{}il}: no iron spread in the globular cluster M22
\footnote{Based on observations collected at the ESO-VLT 
under programs 068.D-0332, 071.D-0217 and 073.D-0211.}}

\author{A. Mucciarelli$^{2}$, E. Lapenna$^{2}$, D. Massari$^{3,4}$,  
E. Pancino$^{3}$, P.B. Stetson$^{5}$,
F.R. Ferraro$^{2}$, \\ 
B. Lanzoni$^{2}$, C. Lardo$^{6}$}
 

\affil{$^{2}$Dipartimento di Fisica \& Astronomia, Universit\`a 
degli Studi di Bologna, Viale Berti Pichat, 6/2 - 40127
Bologna, Italy}

\affil{$^{3}$INAF - Osservatorio Astronomico di Bologna, Via Ranzani 1 - 40127
Bologna, Italy}

\affil{$^{4}$Kepteyn Astronomical Institute, University of Groningen, Landleven 12, 9747 AD Groningen, The Nethelands}

\affil{$^{5}$National Research Council, 5071 West Saanich Road, Victoria, BC V9E 2E7, Canada}

\affil{$^{6}$Astrophysics Research Institute, Liverpool John Moores University, 146 Brownlow Hill, 
Liverpool L3 5RF, United Kingdom}

\begin{abstract}

We present the analysis of high-resolution spectra obtained with UVES and UVES-FLAMES 
at the Very Large Telescope of 17 giants in the globular cluster M22, 
a stellar system suspected to have an intrinsic 
spread in the iron abundance. We find that when surface gravities are derived 
spectroscopically (by imposing to obtain the same iron abundance from FeI and FeII lines) 
the [Fe/H] distribution spans $\sim$0.5 dex, according to previous analyses. 
However, the gravities obtained in this way correspond to
unrealistic low stellar masses (0.1-0.5 $M_{\odot}$) for most of the surveyed giants.
Instead, when photometric gravities are adopted, the [FeII/H] distribution  
shows no evidence of spread at variance with the [FeI/H] distribution. 
This difference has been recently observed in other clusters and
could be due to non-local thermodynamical equilibrium effects driven by 
over-ionization mechanisms, that mainly affect the neutral species (thus providing 
lower [FeI/H]) but leave [FeII/H] unaltered. We confirm that 
the s-process elements show significant star-to-star variations and their abundances 
appear to be correlated with the difference between [FeI/H] and [FeII/H].
This puzzling finding suggests that the peculiar chemical composition of some cluster 
stars may be related to effects able to spuriously decrease [FeI/H].
We conclude that M22 is a globular cluster with no evidence of intrinsic 
iron spread, ruling out that it has retained the supernovae ejecta in its gravitational 
potential well.

\end{abstract}

\section{Introduction}
In the last decade, the investigation of the properties of the globular clusters (GCs) 
and their possible departures from the canonical paradigm of simple stellar population 
has become a hot astrophysical topic. 
The general homogeneity in the Fe content measured in most of them 
\citep[see e.g.][]{carretta09} is considered as the typical feature to distinguish GCs from 
more complex stellar systems \citep{willman12}, suggesting that 
in the past GCs were not massive enough to retain 
the ejecta of supernovae (SNe) in their gravitational well.
On the other hand, the intrinsic star-to-star variations in the elements involved in the 
proton capture processes (the high temperature extension of the CNO-cycle) observed in all 
the old and massive GCs \citep[see e.g.][]{m09,c09,larsen14} suggest that these systems have been 
able to retain low-energy ejecta, possibly synthesized in 
asymptotic giant branch (AGB), fast-rotating massive and/or binary stars 
\citep[see e.g.][]{decressin07,dercole08,demink09,bastian13,denis14}.
However, for a few GC-like systems  
an intrinsic iron spread has been detected, suggesting that those systems were able to retain
the SN ejecta. 
The two undeniable cases are $\omega$ Centauri \citep{origlia03,johnson10,pancino11,marino11a} 
and Terzan~5 \citep{ferraro09,origlia11,origlia13,massari14}, 
both showing a 1 dex wide (and multi-modal) metallicity distribution.
Significant intrinsic iron spreads (smaller than those of $\omega$ Centauri and Terzan 5) 
have been measured from high-resolution spectra in M54 \citep{carretta10a}, 
M22 \citep[][hereafter M09]{marino09}, M2 \citep{yong14} and NGC~5286 \citep{marino15}. 
Other clusters have been proposed to have small intrinsic spreads  but without conclusive results
\citep[see e.g. the case of NGC~1851,][]{carretta10b}.

Indeed, the analysis of GCs suspected to have an intrinsic Fe spread deserves a particular 
care. Recently, the Fe spreads previously measured in two GCs turned out to be spurious and due to 
unaccounted effects. In particular, in NGC~2419 the (spurious) evidence of 
an iron spread measured from the Ca~II triplet 
by \citet{cohen10} and \citet{ibata11} is due to the peculiar chemical composition of the cluster 
(with a strong Mg depletion),
which modifies the strength of the Ca~II triplet lines \citep{mucciarelli12}.
In NGC~3201  the detection of iron spread by \citet{simmerer13} is
due to the inclusion of AGB stars in their sample, having [Fe/H] abundances systematically lower 
than those measured in red giant branch (RGB) stars \citep{m15}. 
The observed effect has been suggested to be due to non-local thermodynamical equilibrium (NLTE) effects that 
affect mainly the neutral species, lowering the abundance 
derived from FeI lines but leaving the abundances from FeII lines unaltered. 
A similar behavior has been observed also in AGB stars
in M5 \citep{ivans01} and 47~Tucanae \citep{lp14}.

Because of these recent results questioning previous claims of iron spread in some GCs, 
here we reanalyse the case of M22.
This metal-poor GC is suspected to have an intrinsic Fe spread since forty years, 
because of the broad colour distribution of RGB observed in its 
color-magnitude diagram (CMD). 
However, the presence of differential reddening in the direction of M22 makes  
difficult to properly assess whether the observed color broadening is caused 
by an intrinsic metal spread \citep{monaco}.
From a spectroscopic point of view, conflicting results have been obtained
based on small samples of stars \citep{cohen,pilachowski,gratton82}.
Recently, M09 and \citet{marino11b} analysed high-resolution 
spectra of 35 giant stars finding that M22 harbors at least two groups of stars characterized 
by different iron, C+N+O and s-process element abundances: the first group has [Fe/H]=--1.82 dex, 
[(C+N+O)/Fe]=+0.28 dex and solar-scaled [s/Fe] abundance ratios, 
while the second group has [Fe/H]=--1.67 dex, 
[(C+N+O)/Fe]=+0.41 dex and [s/Fe]$\sim$+0.3/+0.4 dex. 
Of course, this result puts M22 on a different framework with respect to genuine GCs, 
suggesting that M22 retained not only s-process and CNO-cycle elements (which are typical 
high-mass AGB ejecta) but also the SNe ejecta.

In this paper, we present a new analysis of the sample of 17 giant stars of M22 
that M09 used to provide the first evidence in support to an intrinsic metallicity spread 
in this cluster.



\section{Observations}
The spectroscopic dataset analysed here is the same used by M09 and
includes six giant stars observed 
with UVES@VLT \citep{dekker} on 18-21 March 2002, 
and 11 giant stars observed with UVES-FLAMES@VLT \citep{pasquini} 
on 24-26 May 2003,
adopting the Red Arm 580 grating that ranges from $\sim$4800 to 
$\sim$6800 \AA\  with a typical spectral resolution R=~47000. 
All the spectra have been reduced with the dedicated ESO pipelines, including 
bias subtraction, flat-fielding, wavelength calibration, spectral extraction and 
order merging. The typical signal-to-noise ratio per pixel of the acquired spectra 
is $\sim$150 at $\sim$6000 \AA .

The target stars, originally selected from the photometric catalog by \citet{monaco}, have been 
cross-identified in the UBVI ground-based catalog described in \citet{kunder13} and 
in the $\rm {JHK_s}$ 2MASS catalog \citep{skrutskie}. 
Their position in the (V,B-V) CMD is shown in Fig.~\ref{cmd}.
Main information about the targets is available in M09.

For each target the correction for differential reddening has been derived as in 
\citet{milone12}, adopting the extinction law by \citet{cardelli89}. 
We found that the maximum variation 
of E(B-V) across the area covered by the observed targets is of $\sim$0.07 mag, in nice agreement 
with \citet{monaco} who quoted a maximum variation of $\sim$0.06 mag. 
The true distance modulus, ${\rm (m-M)}_0$=~12.65 mag, and the color excess, E(B-V)=~0.34 mag, of the cluster have 
been estimated by fitting the 
CMD with an isochrone from the BaSTI dataset \citep{pietrinferni06}, 
computed with an age of 12 Gyr, a metallicity Z=~0.0006 and $\alpha$-enhanced chemical mixture 
(corresponding to an iron content of [Fe/H]=--1.84 dex, in agreement with M09). 
The color excess is the same quoted by \citet{harris}, based on the photometry by \citet{cudw}, 
while we derived a slightly fainter ($\sim$0.11 mag) distance modulus. 

\section{Iron abundance}
\label{analysis}

The iron abundances have been derived by comparing observed and theoretical equivalent widths (EWs) 
by means of the code GALA \citep{mucciarelli13}. 
EWs have been measured with the code DAOSPEC \citep{stetson08} run through the wrapper 4DAO \citep{4dao} 
that allows a visual inspection of the best-fit Gaussian profile for each individual line.
Model atmospheres have been computed with the code 
ATLAS9\footnote{http://wwwuser.oats.inaf.it/castelli/sources/atlas9codes.html} 
assuming 1-dimensional, plane-parallel geometry, no overshooting in the computation 
of the convective flux and adopting 
the new opacity distribution functions by \citet{castelli04} computed with an enhanced 
chemical composition for the $\alpha$-elements 
(while for all the other elements a solar [X/Fe] abundance ratio is assumed). 
The metallicity [M/H] of the model atmosphere for each star 
has been chosen according to the average [Fe/H] derived from FeII lines, 
being most of the iron in the ionized stage in the atmosphere of late-type stars.

First guess parameters for effective temperature (\teff) and surface gravities (log~g)
have been calculated from the photometry.
\teff\ has been derived from the color-\teff\ transformations by \citet{alonso99}, 
by averaging the values obtained from different de-reddened broad-band colors, namely
$\rm (U-B)_0$, $\rm (B-V)_0$, $\rm (V-I)_0$, $\rm (V-K_{s})_0$ and $\rm (J-K_{s})_0$.
Surface gravities have been derived through the Stefan-Boltzmann relation, 
assuming the average \teff\ , the bolometric corrections by \citet{alonso99} computed 
with the average \teff\ and the stellar masses obtained from the best-fit isochrone.
For most of the stars we adopted a mass of 0.78 $M_{\odot}$ 
(appropriate for RGB stars, according to the best-fit theoretical isochrone). 
Three targets are identified as likely AGB stars, 
according to their positions in the optical CMDs 
(they are marked as empty triangles in Fig.~\ref{cmd}).
We assumed for these stars a mass of 0.65 $M_{\odot}$, corresponding to the median value of the
mass distribution of the horizontal branch stars of M22 
(obtained by using the zero age horizontal branch models of the BaSTI database). 
The position of two other stars (marked as empty squares in Fig.~\ref{cmd}) could also be compatible 
with the AGB but the small color separation from the RGB makes it difficult to unambiguously assign 
these targets to a given evolutionary sequence.
For these two stars we assume conservatively a mass of 0.78 $M_{\odot}$ and checking that the impact 
of a different mass on the iron abundances is very small: assuming the AGB mass, [FeII/H] 
changes by $\sim$0.03 dex, while [FeI/H] does not change.

Because the targets span relatively large ranges in the parameter space 
($\delta$\teff$\sim$700 K and $\delta$log~g$\sim$1.5 dex, according to the photometric estimates), 
the use of an unique linelist is inadvisable, because the line blending conditions 
vary with the evolutionary stage of the stars.
Hence, a suitable linelist  has been defined for each individual target, by using 
a specific synthetic spectrum calculated with the code SYNTHE \citep[see][for details]{sbordone04}, 
adopting the photometric parameters and including 
only transitions predicted to be unblended and detectable in the observed spectrum.
Each linelist has been refined iteratively: after a first analysis, the selected 
transitions have been checked with synthetic spectra calculated with the new parameters 
and including the precise chemical composition obtained from the analysis. 
The oscillator strengths for FeI lines are from the compilation by \citet{fuhr88} and \citet{fuhr06}, 
while for FeII lines we adopted the recent atomic data by \citet{mel09}. 
Concerning the van der Waals damping constants, the values calculated by \citet{barklem00} are adopted 
whenever possible, while for other transitions they were computed 
according to the prescriptions of \citet{castelli05}. The reference solar value is 7.50 \citep{gs98}.
EW, excitation potential and oscillator strength are listed in Table~1.
The iron abundances have been derived from 130-200 FeI lines and 15-20 FeII lines, 
leading to internal uncertainties arising from the EW measurements 
(estimated as the line-to-line scatter divided to the square root of the number of used lines)
of the order of 0.01 dex (or less) 
for FeI and 0.01-0.02 dex for FeII. The chemical analysis has been performed with
three different approaches to constrain \teff\ and log~g, while the microturbulent velocities 
(\vturb) have been constrained by imposing 
no trend between the iron abundance and the line strength, expressed as $\log(EW/\lambda)$. 
The total uncertainty in the chemical abundance has been computed by summing in quadrature 
the internal uncertainty and that arising from the atmospheric parameters, the latter being estimated 
according to the different method adopted (as discussed below).
Table 2 summarises the average [FeI/H] and [FeII/H] abundances obtained with the different methods.

\subsection{Method (1): spectroscopic \teff\ and log~g}
The values of \teff\ have been derived by erasing any trend between the iron abundance obtained from FeI lines 
 and the excitation potential ($\chi$), while
log~g have been derived by requiring the same abundance from FeI and FeII lines.
Because of the large number of FeI lines, well distributed over a wide range of $\chi$ values, 
the spectroscopic \teff\ are constrained with internal uncertainties of about 30-50 K, while 
the internal uncertainties on log~g are  $\sim$0.03-0.05. 
Uncertainties in \vturb\  are of about 0.1 km/s (this value is valid also for the 
other methods where the same approach is used to derive \vturb). 
We assumed a typical uncertainty of $\pm$0.05 dex in the metallicity [M/H] of the model atmosphere; this 
has a negligible impact on [FeI/H] but leads to variations of $\pm$0.02-0.04 dex in [FeII/H].

The [Fe/H] distributions thus derived from neutral and single ionized lines are 
shown in the left panel of Fig.~\ref{histo1}, as generalized histograms. 
The two distributions are, by construction, very similar to each other (because of the adopted 
constraint to derive log~g) and $\sim$0.5 dex wide, 
with an average value of [Fe/H]=--1.92$\pm$0.03 ($\sigma$=0.13 dex) for both 
[FeI/H] and [FeII/H].
In order to evaluate whether the observed scatter is compatible with an intrinsic spread, 
we adopted the maximum likelihood (ML) algorithm described in \citet{mucciarelli12},
which provides the intrinsic scatter ($\sigma_{int}$) of the metallicity distributions 
by taking into account the uncertainties of each individual star. 
Both the iron distributions have a non-zero scatter, 
with $\sigma_{int}$=0.13$\pm$0.02 dex. 
This result is qualitatively similar to that of M09, who obtained
a broad Fe distribution adopting the same approach to derive the atmospheric parameters.


\subsection{Method (2): spectroscopic \teff\ and photometric log~g}
The values of \teff\ have been constrained spectroscopically, as done in the method (1), 
while those of log~g have been derived through the Stefan-Boltzmann relation. 
In the computation of log~g, we adopted the distance modulus, stellar masses, color excess and 
bolometric corrections used for the guess parameters, together with the spectroscopic \teff.
The internal uncertainty of the photometric log~g has been computed including 
the uncertainties in the adopted \teff, stellar mass, magnitudes and differential reddening corrections, 
leading to a total uncertainty of about 0.05 dex. 
Errors in distance modulus and color excess have been neglected because they impact 
systematically all the stars, while we are interested in the star-to-star uncertainties only.
This approach allows to benefit at best from all the spectroscopic and photometric pieces of information in hand, 
minimizing the impact (mainly on \teff) of the uncertainties in the differential and absolute reddening. 
The atmospheric parameters and the [FeI/H] and [FeII/H] abundance ratios 
derived with this method are listed in Table 3.

By adopting this method, which (at odds with the previous one) does not impose ionization balance, we find 
that, for most of the targets, a large difference between [FeI/H] and [FeII/H].
The [FeI/H] and [FeII/H] distributions are shown in the right panel of Fig.~\ref{histo1}. 
At variance with the previous case, the two distributions look very different: 
the distribution of [FeI/H] spans a range of $\sim$0.5 dex,
with an average value of --1.92$\pm$0.04 ($\sigma$=0.16 dex), while the [FeII/H] distribution 
is narrow and symmetric, with an average value of --1.75$\pm$0.01 dex ($\sigma$=~0.04 dex). 
The ML algorithm provides an intrinsic spread $\sigma_{int}$=0.15$\pm$0.02 dex for the [FeI/H] distribution, 
while the [FeII/H] distribution is compatible 
with a negligible intrinsic scatter ($\sigma_{int}$=~0.00$\pm$0.02 dex).

To illustrate this difference between [FeI/H] and [FeII/H], 
Fig.~\ref{spec2} shows some FeI and FeII lines in the spectra of stars \#200080
(where [FeI/H] is 0.29 dex lower than [FeII/H]) and \#88 (where 
[FeI/H] and [FeII/H] differ by 0.05 dex only). In the first case, 
the synthetic spectrum calculated with the average abundance derived from FeII lines 
(red solid line) is not able to reproduce the FeI lines.
The latter are always weaker than those of the synthetic spectrum, 
regardless of their $\chi$ and line strength, thus suggesting that the discrepancy 
is not due to inaccuracies in \teff\ and/or \vturb\  (otherwise a 
better agreement would have been found for high-$\chi$ lines, less sensitive to \teff, 
and/or for weak lines, less sensitive to \vturb).
On the other hand, the synthetic spectrum computed with the [FeI/H] abundance 
(blue dashed line) does not fit the FeII line, 
that is stronger than that predicted by the synthetic spectrum.
In the case of star \#88 the situation is different and 
an unique Fe abundance is able to well reproduce both FeI and FeII lines.

\subsection{Method (3): photometric \teff\ and log~g} 
As an additional check, the analysis has been performed keeping  \teff\ and log~g fixed
at the guess values derived from the photometry (see Section 3), and optimizing spectroscopically only \vturb. 
This set of parameters  is very similar to that obtained with 
method (2), with the average differences, in the sense of method (3) - method (2), 
of +58$\pm$12 K ($\sigma$=~50 K) in \teff, +0.02$\pm$0.005 ($\sigma$=0.02) 
in log~g and +0.04$\pm$0.02 km$s^{-1}$ ($\sigma$=~0.08 km$s^{-1}$) in \vturb . 
In particular, we note that spectroscopic and photometric \teff\ agree very well, and their 
differences do not show trends with the photometric \teff , as visible in Fig.~\ref{difft} where 
the difference between \teff\ from method (3) and (2) are shown as a function of the spectroscopic \teff.

Fig.~\ref{histo2} shows the Fe abundance distributions obtained with 
the average photometric parameters (upper-left panel) and using
the individual broad-band colors
$\rm (U-B)_0$, $\rm (B-V)_0$, $\rm (V-I)_0$, $\rm (V-K_{s})_0$, $\rm (J-K_{s})_0$. 
In all cases, the [FeII/H] distribution is single-peaked and narrow, well 
consistent with that obtained from method (2). Instead, 
whatever color is adopted, the [FeI/H] distribution always has a much larger (by a factor of 2-3) 
dispersion than that obtained for [FeII/H], similar to the 
finding of method (2). 
In particular, when we consider the average photometric parameters, 
the ML algorithm provides intrinsic scatter of 0.12$\pm$0.02 for [FeI/H] and 
0.00$\pm$0.02 for [FeII/H].
Because the results obtained with this method agree with those obtained with method (2), 
and the star-to-star uncertainties in spectroscopic \teff\ are smaller than 
the photometric ones (which are also affected by the uncertainties on the differential reddening corrections),
in the following we refer only to method (2) as alternative approach to method (1).

\section{A sanity check: NGC~6752}
\label{refe}
As a sanity check, UVES-FLAMES archival spectra of 14 RGB stars in the GC NGC~6752 
observed with the Red Arm 580 grating have been analysed following the same  
procedure used for M22.
NGC~6752 is a well-studied GC that can be considered as a standard example of {\sl genuine} 
GC, with no intrinsic iron spread \citep[see e.g.][]{yong05,carretta09}\footnote{ \citet{yong13} performed 
a strictly differential line-by-line analysis on 37 RGB stars of NGC~6752 by using high-quality UVES spectra, 
finding an observed spread in [Fe/H], 0.02 dex, larger of a factor of 2 than the internal uncertainties.
This small intrinsic spread could reflect He variations and/or real inhomogeneities in the cluster iron content. 
Because such chemical inhomogeneities can be revealed only when the internal uncertainties are smaller than 
$\sim$0.02 dex, for our purposes we can consider NGC~6752 as a {\sl genuine} GC.}
and with a metallicity comparable with that of M22.
This approach allows to remove any systematics due to the adopted atomic data, solar 
reference values, model atmospheres, method to measure EWs and to derive the atmospheric 
parameters.
When the parameters are derived following method (2),
we derive average abundances [FeI/H]=--1.62$\pm$0.01 dex ($\sigma$=~0.04 dex) and 
[FeII/H]=--1.58$\pm$0.01 dex ($\sigma$=~0.04 dex), in good agreement with the previous 
estimates available in the literature.
In this case, the two iron distributions (shown in Fig.~\ref{6752md}) have small observed dispersions, 
both compatible with a negligible scatter within the uncertainties, as demonstrated by the ML algorithm. 
The two distributions are compatible with each other also in terms of their shape, 
at variance with those of M22. 
The same results are obtained when the parameters are all derived spectroscopically.
This test demonstrates that: (i)~the different [FeI/H] and [FeII/H] distributions obtained 
for M22 with methods (2) and (3) are not due to the adopted procedure; 
(ii)~in a {\rm normal} GC the shape of [FeI/H] and [FeII/H] distributions are not significantly different.

\section{No iron spread in M22}
The new analysis of the sample of giant stars already discussed in M09 leads to an unexpected result:
an iron abundance spread in M22 is found when FeI lines are 
used, independently of the adopted spectroscopic or photometric gravities.
{\sl This scatter totally vanishes when the iron abundance is derived from FeII lines 
and photometric gravities are used.} In the case of spectroscopic gravities, 
the abundances from FeII lines are forced to match those from FeI lines, thus 
producing a broad [FeII/H] distribution. Given that the adoption of 
photometric gravities leads to a broad [FeI/H] distribution and a narrow, mono-metallic [FeII/H] distribution,
which one should we trust?
In principle, FeII lines are most trustworthy than FeI lines to determine the iron abundance, 
because FeII is a dominant species in the atmospheres of late-type stars 
(where iron is almost completely ionized) and its lines 
are unaffected by NLTE effects, at variance with the FeI lines 
\citep[see e.g.][]{kraft03,mashonkina11}.

The analysis of the results shown in Fig.~\ref{histo1} and \ref{histo2} suggests that the adoption of method (1) 
tends to produce an artificial spread of [FeII/H] toward low metallicities. 
Since [FeII/H] strongly depends on the adopted values of logg, this implies that gravities are severly 
underestimated in method (1). 
This bias  is clearly revealed when the stellar masses corresponding to the spectroscopic values of
log~g values are computed.
We estimated the stellar masses by inverting the Stefan-Boltzmann equation and assuming the spectroscopic 
log~g derived with method (1). 
The derived masses range from 0.12 to 0.79 $M_{\odot}$, with a 
mean value of 0.46 $M_{\odot}$ and a dispersion of 0.2 $M_{\odot}$. 
Note that $\sim$70\% of the stars have masses below 0.6 $M_{\odot}$. Such low values, 
as well as the large dispersion of the mass distribution, are unlikely for a sample 
dominated by RGB stars, with expected masses close to 0.75-0.80 $M_{\odot}$. In particular, 
10 target stars have log~g that would require masses below 0.5 $M_{\odot}$, 
thus smaller than the typical mass of the He-core of GC giant stars at the luminosity level 
of our targets. Such very low masses cannot be justified even in light of the uncertainties 
in the mass loss rate \citep{origlia14}. A similarly wide mass distribution is obtained 
by adopting the spectroscopic parameters by M09, leading to a mass range between 0.34 and 1.19 $M_{\odot}$.
In that case, three stars have masses larger than 0.8 $M_{\odot}$, corresponding to 
the typical mass of a turnoff star of M22. 
For comparison, the masses derived from the spectroscopic log~g of the spectral sample 
of NGC~6752 (see Section~\ref{refe}) cover a small and well reasonable range, from 0.65 to 0.85 $M_{\odot}$, 
with an average value of 0.75 $M_{\odot}$ ($\sigma$=0.06 $M_{\odot}$).

Fig.~\ref{mass} shows 
the behavior of the difference [FeI/H]-[FeII/H], as derived with method (2), as a function 
of the stellar masses, as derived from the spectroscopic gravities in method (1). 
The mass intervals expected for RGB and AGB stars 
in the luminosity range of our spectroscopic targets are shown as grey shaded regions. 
A clear trend between the [FeI/H]-[FeII/H] difference and the stellar mass is found.
The stars with the largest difference between FeI and FeII abundances are also those 
where the spectroscopic log~g requires an unrealistically low mass, while for the stars where 
[FeI/H] is consistent with [FeII/H] the spectroscopic log~g provide masses in reasonable agreement 
with the theoretical expectations.
This demonstrates that the spectroscopic gravities needed to 
force [FeII/H] matching the low-abundance tail of the [FeI/H] distribution
lead to unreliable stellar masses.
Since this is the only case in which [Fe II/H] shows significant
spread, we have to conclude that the observed large iron distribution is 
not real. The correct diagnostic of iron content therefore are the
Fe II lines analyzed under the assumption of photometric
gravities. These always lead to a narrow iron distribution (see
Figs.~\ref{histo1} and \ref{histo2}), thus implying that no iron spread is observed in M22.


An additional confirmation of the different behavior of neutral and ionized lines in our sample is provided by 
the analysis of the titanium transitions, because this element is one of the few species that provides 
a large number of both neutral and single ionized lines. 
The oscillator strengths are from \citet{mfw} and \citet{lawler13} for Ti~I lines and 
from \citet{wood13} for Ti~II lines. The [TiI/H] and [TiII/H] abundances exhibit 
the same behavior discussed above for the Fe abundances. When the spectroscopic gravities 
are used, both the distributions are broad, with an observed scatter of $\sim$0.2 dex (see left panel of Fig.~\ref{tid}). 
On the other hand, when the photometric gravities are adopted (see Table 3), the [TiII/H] distribution is consistent 
with null intrinsic scatters, while that of [TiI/H] remains broad and skewed toward low abundances 
(right panel of Fig.~\ref{tid}).
We note that the difference [TiI/H]-[TiII/H] strongly correlates with the difference 
[FeI/H]-[FeII/H], with a Spearman rank correlation coefficient $C_S$=+0.956 that provides 
a probability of $\sim10^{-8}$ that the two quantities are not correlated.
Hence, the analysis of [TiI/H] and [TiII/H] reinforces the scenario where the abundances from 
neutral lines in most of the M22 stars 
are biased, providing distributions (artificially) larger than those from single ionized lines. 

\section{The s-process elements abundance}
M09 and \citet{marino11b} found that M22 has, together with a dispersion in the iron content, 
an intrinsic spread in the abundances of s-process elements. In light of the results described above, 
we derived abundances also for these 
elements, by adopting the parameters obtained with method (2) and measuring  
Y~II, Ba~II, La~II and Nd~II lines.
For Y and Nd the abundances have been obtained with GALA from the EW measurement, 
as done for the Fe and Ti lines, and adopting the oscillator strengths available in the 
Kurucz/Castelli linelist. Ba~II and La~II lines are affected by hyperfine and isotopic 
splittings. 
The linelists for the La~II lines are from \citet{lawler01}, while those for 
the Ba~II lines from the NIST database \footnote{http://physics.nist.gov/PhysRefData/ASD/lines\_form.html}.
Only for these two elements, the abundances have been derived 
with our own code SALVADOR (A. Mucciarelli et al. in preparation) that performs a
$\chi^{2}$-minimization between observed and synthetic spectra calculated with 
the code SYNTHE.

For all these elements, we found that the absolute abundances show large star-to-star variations, 
with observed scatters between $\sim$0.2 and $\sim$0.3 dex, depending on the element. These spreads 
are not compatible within the uncertainties.
Because of the possible occurrence of the NLTE effects, 
the abundance ratios [X/Fe] (see Table 4)
have been estimated by using the FeII abundances as reference; in fact, for these elements 
the chemical abundances have been derived only from single ionized transitions,
which are less sensitive to the overionization 
\citep[or sensitive to it in a comparable way to the FeII lines; see e.g. the discussion in][]{ivans01}. 
The [X/FeII] abundance ratios show significant intrinsic spreads, 
as confirmed by the ML algorithm.
Note that, if we adopt FeI abundances as reference, the [X/FeI] abundance ratios 
still display an intrinsic scatter, because the observed spread in the absolute 
abundances for these s-process elements is larger than that measured from the FeI lines.

Fig.~\ref{spro1} shows the behavior of each s-process element abundance ratio 
as a function of the difference between [FeI/H] and [FeII/H]. 
In all the cases, a clear trend between [X/FeII] and [FeI/H]-[FeII/H] 
is detected, in the sense that the stars characterized by higher s-process abundances 
display a better agreement between FeI and FeII. In the case of Y, Ba and Nd, we find two distinct and 
well separated groups of stars, while for La the behavior is  continuous, 
with no clear gap.
Finally, Fig.~\ref{spro2} plots the behavior of $<$[s/FeII]$>$, obtained by averaging 
together the four abundance ratios, as a function of [FeI/H]-[FeII/H], confirming the existence 
of two groups of stars, with different [s/Fe] and [FeI/H] (but the same [FeII/H]). 
This finding resembles the results by M09 who identify two groups
of stars, named {\sl s-poor} and {\sl s-rich}.


\section{Discussion: re-thinking M22}

\begin{figure}
\plotone{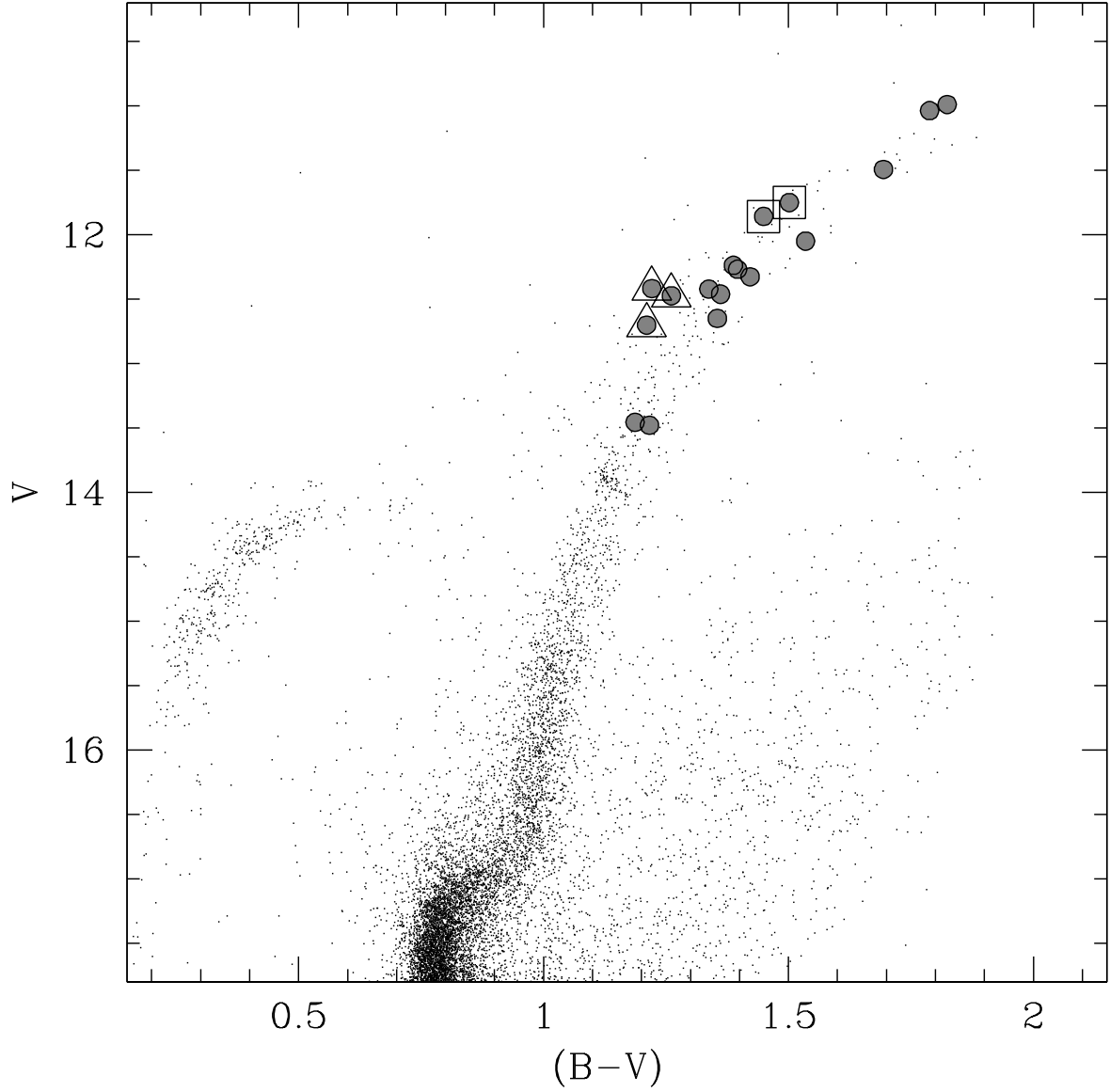}
\caption{(B-V, V) color-magnitude diagram of M22 \citep{kunder13} with marked as grey circles the spectroscopic targets. 
Empty triangles are the likely candidate AGB stars, while empty squares are possible (but not sure) AGB stars.}
\label{cmd}
\end{figure}

\begin{figure}
\plotone{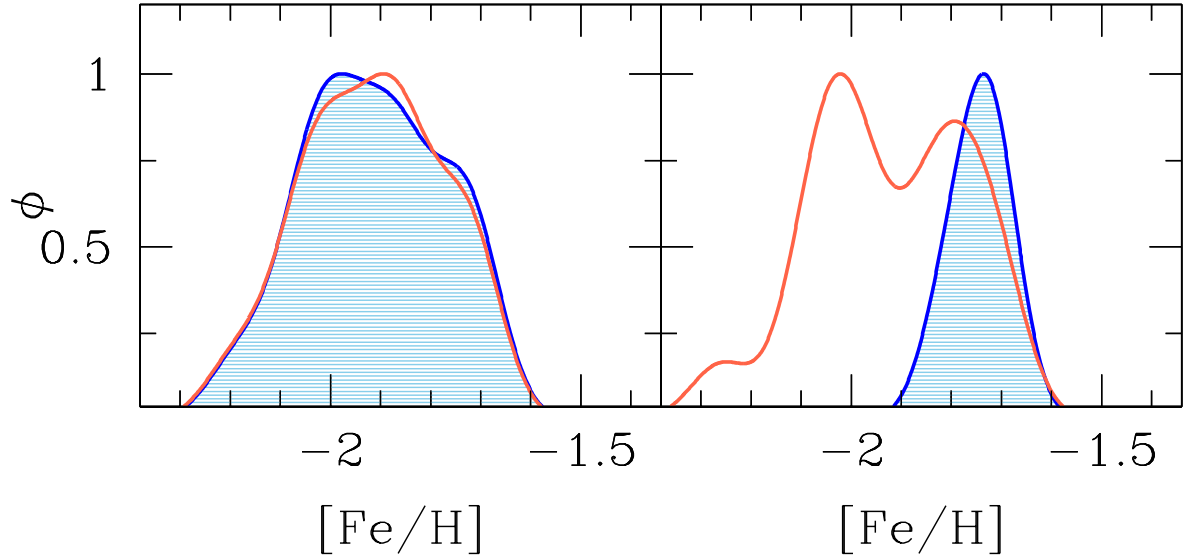}
\caption{Generalized histograms for [FeI/H] (empty red histogram) and [FeII/H] (blue histogram) obtained 
from the analysis performed with spectroscopic gravities (method (1), left panel) and with photometric 
gravities (method (2), right panel).}
\label{histo1}
\end{figure}

\begin{figure*}
\includegraphics[angle=-90,scale=0.65]{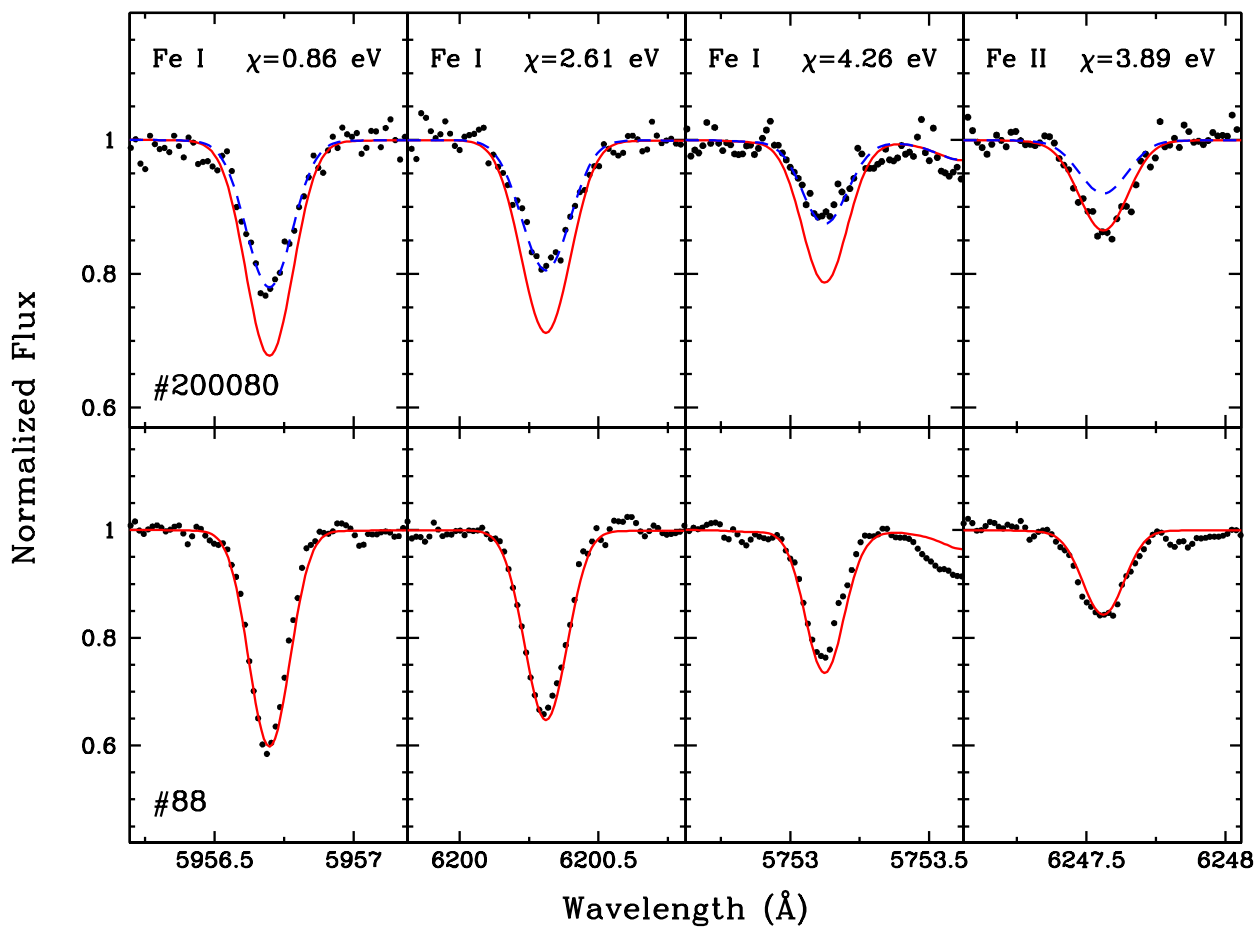}
\caption{Spectral regions around three Fe I lines with different excitation potential and 
one Fe II line, for the target stars \#200080 (upper panels) and \#88 (lower panels). 
Synthetic spectra calculated with the corresponding atmospheric parameters (see Table 1) 
and adopting the average iron abundance derived from Fe II lines are superimposed as red curves. 
The blue dashed curve shown in the upper panels 
is the synthetic spectrum calculated with the iron abundance derived from Fe I lines.}
\label{spec2}
\end{figure*}

\begin{figure}
\plotone{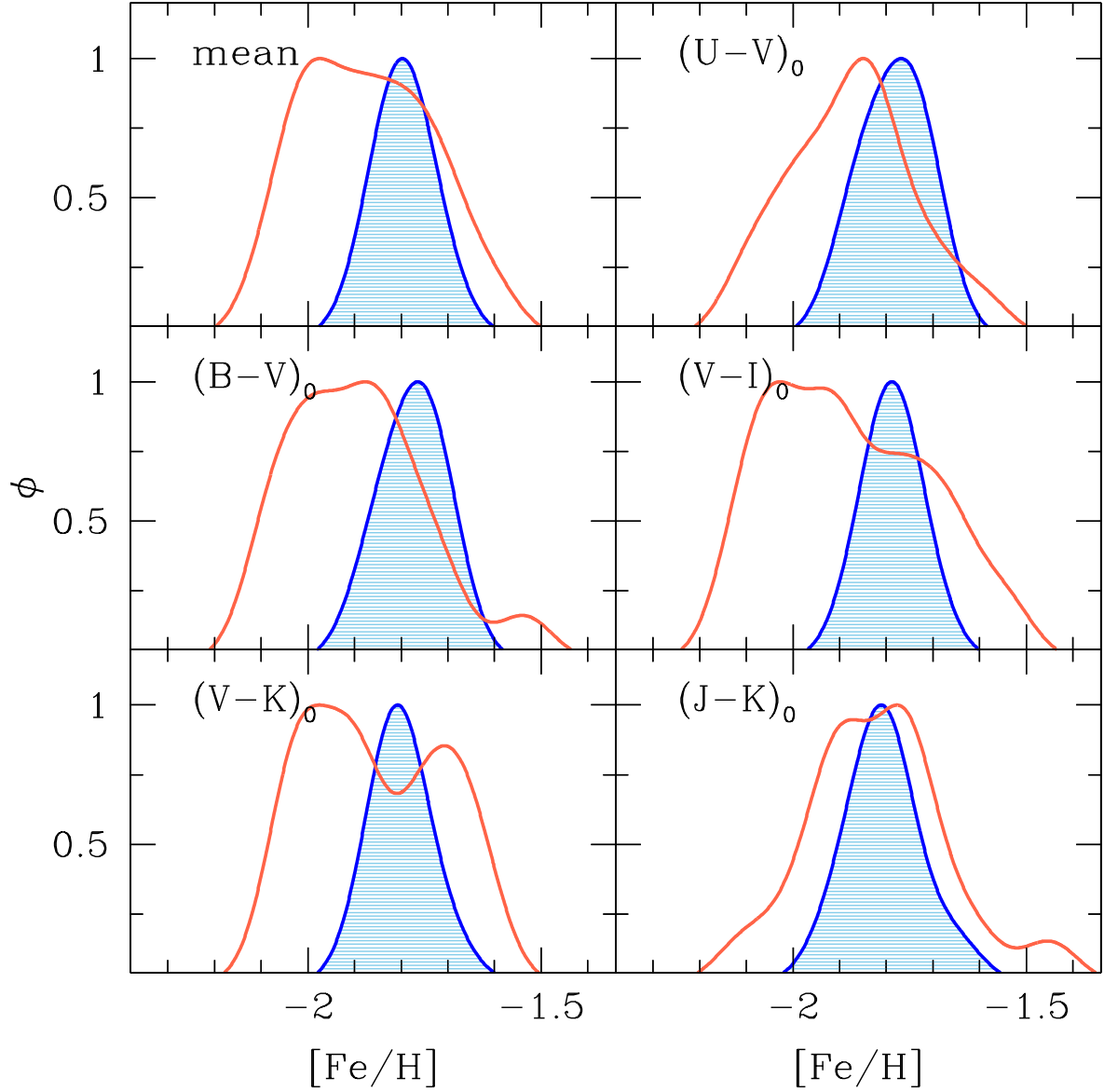}
\caption{Generalized histograms for [FeI/H] and [FeII/H] (same colors of Fig.~\ref{histo1}) obtained 
with the method (3) (photometric \teff\ and log~g), adopting the mean parameters (left-upper panel) and 
those derived from individual broad-band colors. }
\label{histo2}
\end{figure}

\begin{figure}
\plotone{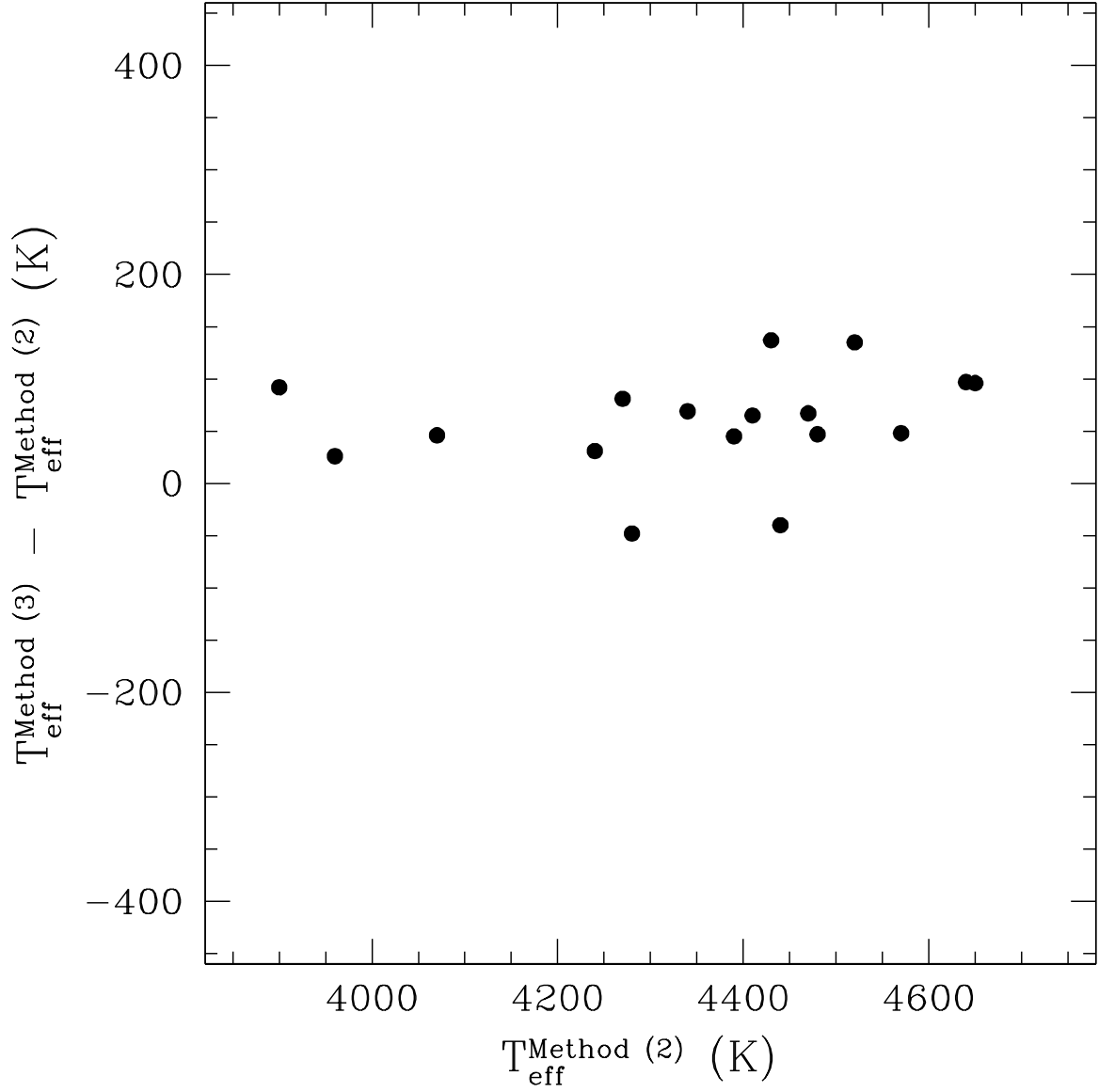}
\caption{ Behaviour of the difference between \teff\ as derived with method (3) and (2) as a function of those 
derived with method (2).}
\label{difft}
\end{figure}

\begin{figure}
\plotone{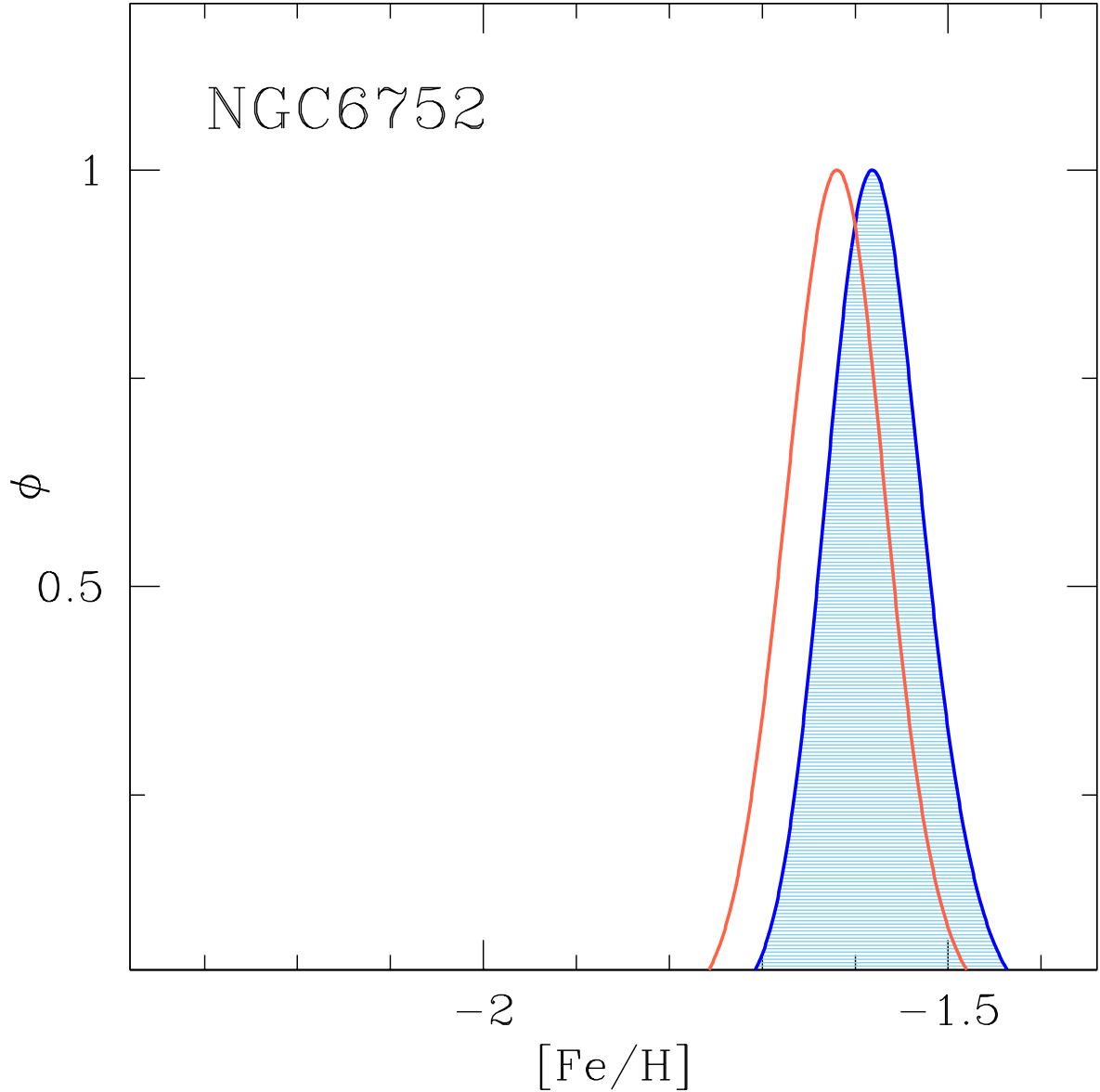}
\caption{Generalized histograms for [FeI/H] and [FeII/H] (same colors of Fig.~\ref{histo1}) 
for a sample of 14 RGB stars in the GC NGC~6752. The analysis has been performed adopting 
spectroscopic \teff\ and photometric log~g, the same method used for the right panel of Fig.~\ref{histo1}.}
\label{6752md}
\end{figure}

\begin{figure*}
\plotone{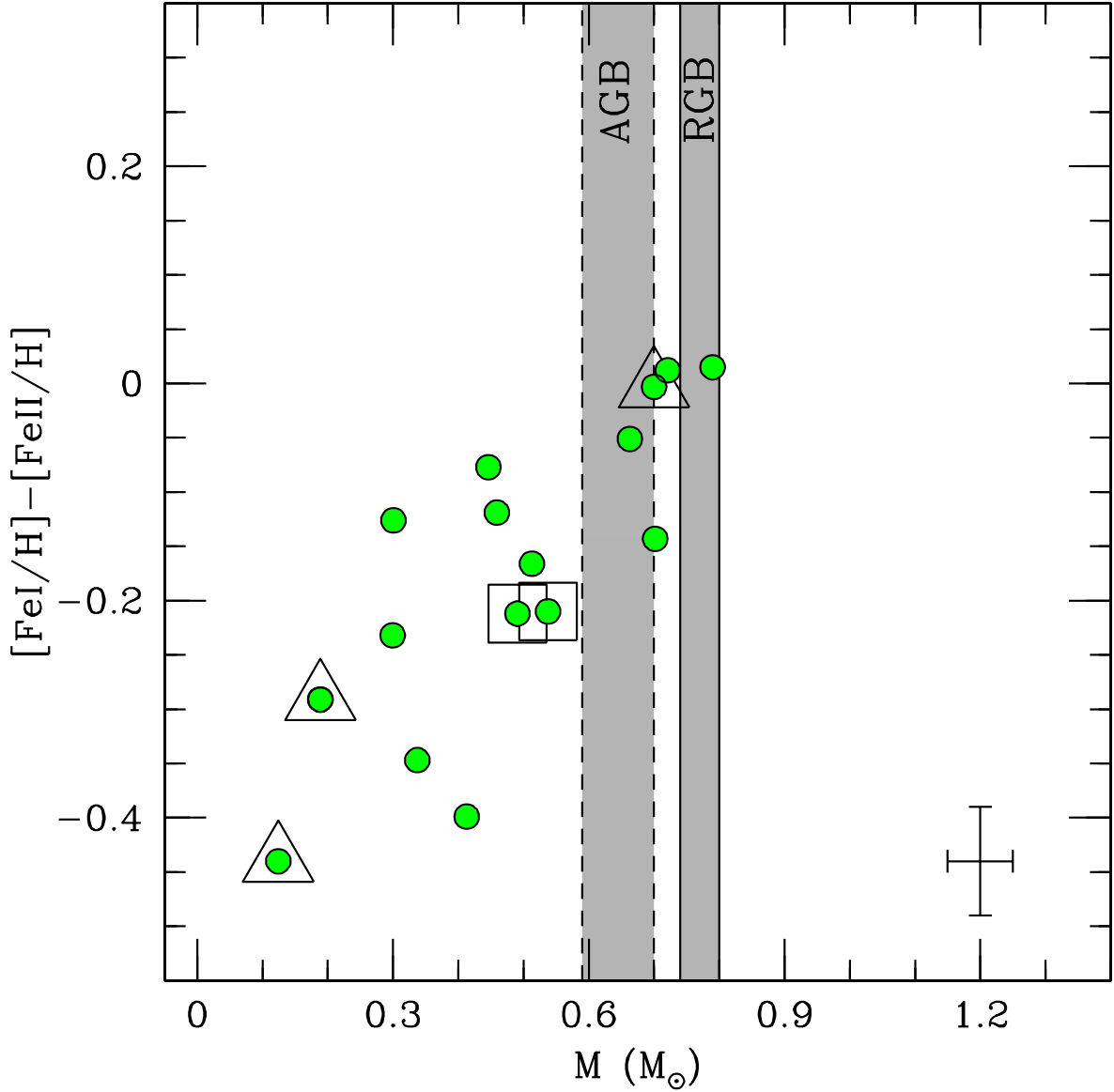}
\caption{Behavior of the difference [FeI/H]-[FeII/H], as derived with method (2), of the spectroscopic 
targets as a function of the stellar masses inferred from the spectroscopic log~g in method (1). 
The two shaded grey regions mark the mass range expected for AGB and RGB stars. 
Same symbols of Fig.~1.}
\label{mass}
\end{figure*}

\begin{figure}
\plotone{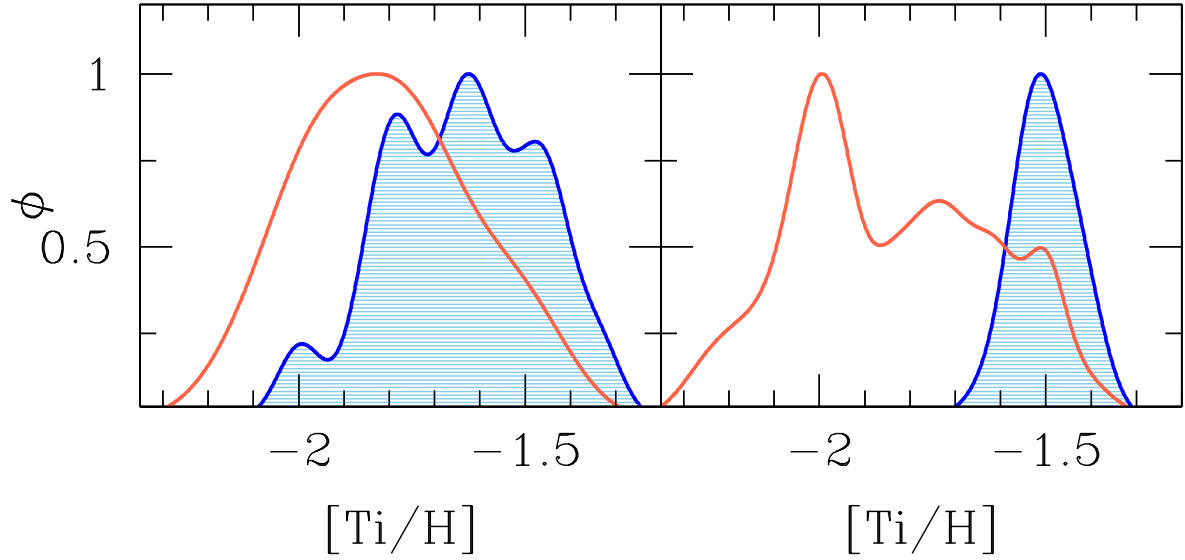}
\caption{Generalized histograms for [TiI/H] (empty red histogram) and [TiII/H] (blue histogram) obtained adopting the
spectroscopic (left panel) and photometric log~g (right panel).}
\label{tid}
\end{figure}

\begin{figure*}
\plotone{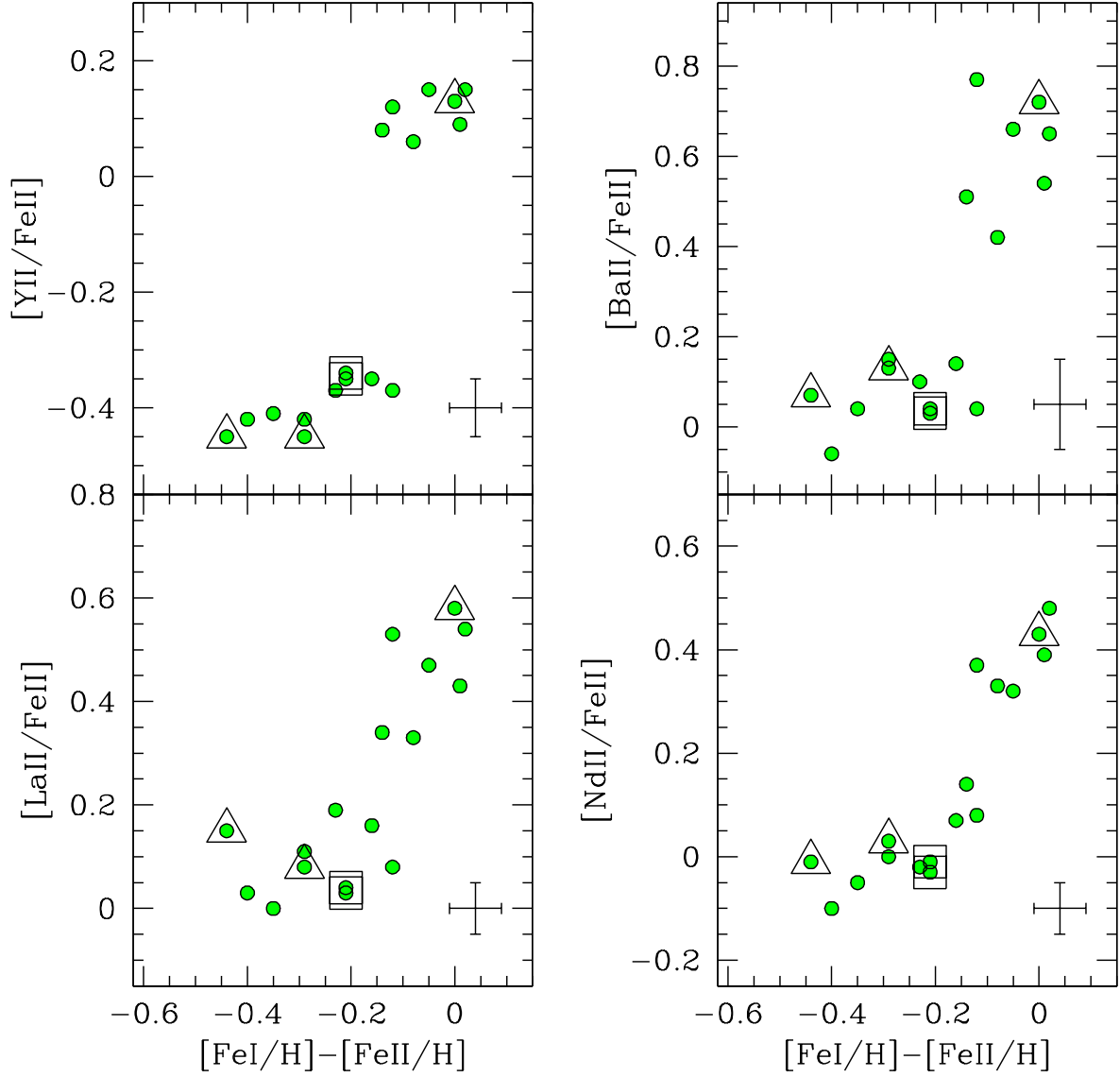}
\caption{Behavior of the abundance of the s-process elements 
Y, La, Ba and Nd as a function of the difference 
between [FeI/H] and [FeII/H]. Same symbols of Fig.~1.}
\label{spro1}
\end{figure*}

\begin{figure*}
\plotone{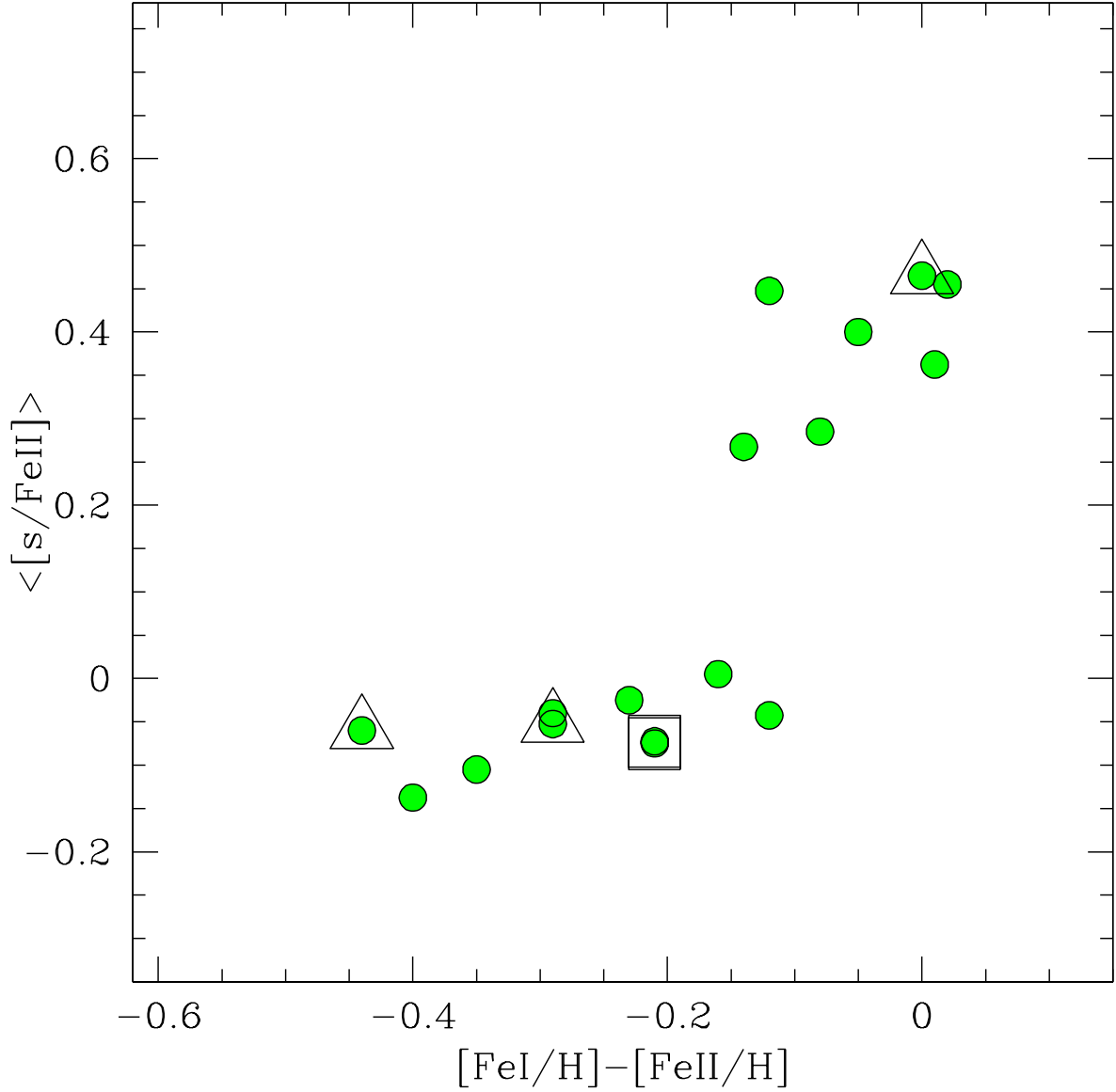}
\caption{Behavior of the average abundance of s-process elements 
(derived by averaging together the abundances of Y, La, Ba and Nd)
 as a function of the difference 
between [FeI/H] and [FeII/H]. Same symbols of Fig.~1.}
\label{spro2}
\end{figure*}

The main results and conclusions of this work are summarized as follows:
\begin{itemize}
\item 
The new analysis of M22 presented here demonstrates that this GC is mono-metallic and that 
the previous claim of a  metallicity scatter was due to a systematic under-estimate of the FeI abundance 
combined with the use of spectroscopic gravities.
When photometric log~g are adopted, the FeII lines 
provide the same abundance for all the stars, regardless of the adopted method to estimate \teff . 
\item 
In light of this result, the formation/evolution scenario for M22 must be 
deeply re-thought.  The homogeneity in its iron content 
suggests that M22 was not able to retain the SN ejecta in its gravitational well. 
Hence, it is not necessary to invoke that the cluster was significantly more massive at its birth
and that it subsequently lost a large amount of its mass.
The observed unimodal [FeII/H] distribution rules out 
the possibility that M22 is the remnant of a now disrupted dwarf galaxy, 
because these systems are characterized by a wide range of metallicity, due to
the prolonged star-formation activity \citep[see][and references therein]{tolstoy09}. 
Also, comparisons between M22 and $\omega$ Centauri \citep{dcm11} are 
undermined by the homogeneity in the [FeII/H] abundance of M22. 
On the other hand, M22 cannot be considered as a {\sl genuine} GC, because of
the intrinsic spread in heavy s-process elements abundances, pointing out the occurence of 
a peculiar chemical enrichment (probably from AGB stars) in this cluster, at variance with 
most of the GCs where s-process elements do not show intrinsic scatters 
\citep{dorazi10}.
\item
M09 and \citet{marino11b} discussed the possibility that M22 is the product of a merging between two 
GCs with different chemical composition. In light of our new analysis, 
this scenario appears unlikely, even if it cannot be totally ruled out.
In this framework, M22 should form from the merging between two clusters with 
the same Fe content, but characterized by different s-process element abundances. 
While clusters with comparable metallicity and different s-process abundance are indeed observed 
\citep[for instance  M4 and M5;][]{ivans99,ivans01}, in this scenario
the cluster with normal s-process abundances should be composed mainly by stars with a large difference 
between [FeI/H] and [FeII/H], while the second cluster should have stars with enhanced s-process abundances 
and similar [FeI/H] and [FeII/H] (see Fig.~\ref{spro2}).
\item As a possible working hypothesis to explain the observed behavior of [FeI/H] and [FeII/H], 
we note that the difference between [FeI/H] and [FeII/H]
is qualitatively compatible with the occurrence of NLTE
effects driven by overionization. These effects are known to affect mainly the less abundant species,  
like FeI, and to have a negligible/null impact on the dominant species, like FeII 
\citep[see e.g.][]{thevenin99,mashonkina11,fabrizio12}. 
Under NLTE conditions, the spectral lines of neutral ions are weaker than in LTE. 
Hence, when the line formation is calculated in LTE conditions (as done in standard analyses), 
the resulting abundance of neutral lines will be correspondingly lower.\\
The same interpretative scheme can be applied to M22. A large and intrinsically broad 
Fe distribution is obtained only from FeI lines, according to the systematic 
underestimate of the Fe abundance obtained when lines affected by overionization are 
analysed in LTE. On the other hand, FeII lines are not affected by NLTE and they 
provide (when photometric log~g are used) the correct abundance, leading to a 
narrow abundance distribution.
\item The mismatch between [FeI/H] and [FeII/H] observed in M22 resembles those found in the GCs
M5 \citep{ivans01}, 47~Tucanae \citep{lp14} and NGC~3201 \citep{m15}.
In these cases, the different behavior observed for FeI and FeII lines 
is restricted to AGB stars only, where FeI lines provide abundances systematically 
lower than those from FeII lines,
while RGB stars have similar [FeI/H] and [FeII/H]. 
However, the situation  is more complex in M22, because a large difference 
between [FeI/H] and [FeII/H] is observed in most of the stars and not only in AGB stars.
Among the target stars of M22, three are identified as AGB stars, 
according to their positions in the CMDs. 
Two of them have a large Fe difference, [FeI/H]-[FeII/H]=--0.29 and --0.44 dex, while 
for the third star FeI and FeII lines provide almost the same abundance. 
The other two possible AGB stars (empty squares in Fig.~\ref{cmd}) have Fe differences of 
--0.21 dex. On the other hand, comparable differences are observed among some RGB stars. 
For instance, the two faintest stars of the sample (\#221 and \#224) are clearly RGB stars 
(see Fig.~\ref{cmd} and \ref{mass}), because they are located at the luminosity level where the color separation between 
RGB and AGB is the largest. On the other hand, these two stars (with very similar atmospheric 
parameters and [FeII/H]) have different [FeI/H] abundances: 
star \#224 has a difference of [FeI/H]-[FeII/H]=--0.14 dex, while star \#221 has [FeI/H]-[FeII/H]=--0.29 dex. 
If departures from LTE are the reasons for the observed discrepancy between [FeI/H] and [FeII/H]  (and between [TiI/H] and [TiII/H]), 
this finding challenges the available NLTE calculations \citep[see e.g.][]{lind12,bergemann}, in which
stars with very similar parameters are expected to have the same NLTE corrections. 
New NLTE calculations should be performed to investigate this hypothesis, with the constraint 
to reproduce simultaneously the discrepancies in Fe and Ti.

\item We found that the difference between [FeI/H] and [FeII/H] is correlated with the s-process element 
abundances.
The behavior is quite puzzling, because the stars with an anomalous difference between [FeI/H] and 
[FeII/H] are those with {\sl normal} s-process abundances, compatible with the abundances observed 
in most of the GCs and in the Galactic field stars of similar metallicity \citep[see e.g. Fig.3 in ][]{venn04}. 
On the other hand, the stars 
enriched in s-process elements show a good agreement between [FeI/H] and [FeII/H].
{\sl Whatever the mechanism responsible to spread [FeI/H] is, it must be also responsible for the 
peculiar behavior of the s-process element abundances.}

\item We confirm the claim already suggested by 
\citet{lp14} and \citet{m15}: chemical analyses based on 
FeI lines and spectroscopic gravities can lead to spurious abundance spreads. 
In light of these results, any claim of intrinsic iron spread in GCs should be always confirmed with an analysis 
based on FeII lines and photometric gravities. If the abundance spread is real,  
it should be detected also when FeII lines and photometric log~g are adopted, 
since FeII lines are the most reliable indicators of the iron abundance. 
All the GCs with anomalous intrinsic Fe spreads observed so far \citep[see][for an updated list]{marino15} 
deserve new analyses in light of this effect, in order to firmly establish whether these spreads are 
real or spurious. 
\end{itemize}

\acknowledgements  
We warmly thank the anonymous referee for suggestions in improving the paper. 
We kindly acknowledge P. Bonifacio, B. Dias, R. Gratton, L. Mashonkina, T. Merle and  L. Monaco for 
comments, helps and suggestions.  
This research is part of the project COSMIC-LAB (http://www.cosmic-lab.eu) funded 
by the European Research Council (under contract ERC-2010-AdG-267675).


\begin{deluxetable}{lccccc}
\tablecolumns{6} 
\tablewidth{0pc}  
\tablecaption{Star identification number, wavelength, ion, excitation potential, oscillator strength 
and measured EWs for all the used transitions.}
\tablehead{ 
\colhead{Star} &  $\lambda$ & Ion & $\chi$ & log~gf & EW \\
               &     (\AA)  &         & (eV)   &        & (m\AA)}
\startdata 
\hline 
  51 & 4805.415   &    TiI	& 2.340   &   0.070   &   30.4   \\  
  51 & 4870.126   &    TiI	& 2.250   &   0.440   &   46.7    \\ 
  51 & 4885.079   &    TiI	& 1.890   &   0.410   &   76.1    \\ 
  51 & 4909.098   &    TiI	& 0.830   &  -2.370   &   15.8    \\ 
  51 & 4913.614   &    TiI	& 1.870   &   0.220   &   64.0     \\	    
  51 & 4915.229   &    TiI	& 1.890   &  -0.910   &   11.8    \\ 
  51 & 4919.860   &    TiI	& 2.160   &  -0.120   &   30.1    \\ 
  51 & 4926.148   &    TiI	& 0.820   &  -2.090   &   20.9   \\ 
  51 & 4937.726   &    TiI	& 0.810   &  -2.080   &   27.2   \\  
  51 & 4997.096   &    TiI	& 0.000   &  -2.070   &   90.7   \\  
  51 & 5009.645   &    TiI	& 0.020   &  -2.200   &   83.1   \\  
  51 & 5016.161   &    TiI	& 0.850   &  -0.480   &  107.5   \\  
  51 & 5020.026   &    TiI	& 0.840   &  -0.330   &  117.0   \\	
  51 & 5036.464   &    TiI	& 1.440   &   0.140   &   98.9   \\	
  51 & 5038.397   &    TiI	& 1.430   &   0.020   &   95.9  \\  
  51 & 5043.584   &    TiI	& 0.840   &  -1.590   &   46.6   \\  
  51 & 5045.415   &    TiI	& 0.850   &  -1.840   &   29.7   \\	
  51 & 5052.870   &    TiI	& 2.170   &  -0.270   &   23.0   \\	
  51 & 5062.103   &    TiI	& 2.160   &  -0.390   &   15.6  \\  
  51 & 5065.985   &    TiI	& 1.440   &  -0.970   &   38.1   \\  
\hline
 \hline
\enddata 
\tablecomments{This table is available in its entirety in machine-readable form.}
\end{deluxetable}

\begin{deluxetable}{lcccccc}
\tablecolumns{7} 
\tablewidth{0pc}  
\tablecaption{Observed and intrinsic scatters for [FeI/H] and [FeII/H] as derived from the ML algorithm and from the
three methods described in the paper.}
\tablehead{ 
 & [FeI/H] & $\sigma_{obs}$ & $\sigma_{int}$  & [FeII/H] & $\sigma_{obs}$ & $\sigma_{int}$  \\
 &  &  &  & & & }
\startdata 
\hline 
 {\rm Method 1}      &  --1.92$\pm$0.03   &  0.14 & 0.13$\pm$0.02  &  --1.90$\pm$0.03 & 0.14  & 0.13$\pm$0.02 \\
 {\rm Method 2}      &  --1.92$\pm$0.04   &  0.16 & 0.15$\pm$0.02  &  --1.75$\pm$0.01 & 0.04  & 0.00$\pm$0.02 \\
 {\rm Method 3}      &  --1.86$\pm$0.03   &  0.13 & 0.12$\pm$0.02  &  --1.81$\pm$0.01 & 0.05  & 0.00$\pm$0.02 \\
\hline
 \hline
\enddata 
\end{deluxetable}


\begin{deluxetable}{lcccccccc}
\tablecolumns{9} 
\tablewidth{0pc}  
\tablecaption{ Atmospheric parameters, [FeI/H], [FeII/H], [TiI/H] and [TiII/H] abundances for the 
spectroscopic targets of M22, as derived with method (1). 
The last line lists the average abundances with the statistical error.}
\tablehead{ 
\colhead{Star} &   \teff\ & log~g & \vturb\ & [FeI/H] & [FeII/H] & [TiI/H] &  [TiII/H] & Notes\\
  &   (K)  &    &(km/s) & &  &  &  &}
\startdata 
\hline 
      51  &  4280  &	1.00	   &  1.70   &  --1.70$\pm$0.02   &   --1.71$\pm$0.04	&    --1.50$\pm$0.04  &  --1.34$\pm$0.04  &   \\ 
      61  &  4430  &	0.95	   &  1.70   &  --1.85$\pm$0.05   &   --1.84$\pm$0.04	&    --1.74$\pm$0.06  &  --1.61$\pm$0.04  &   \\ 
      71  &  4405  &	0.97	   &  1.50   &  --1.90$\pm$0.04   &   --1.89$\pm$0.04	&    --1.77$\pm$0.04  &  --1.59$\pm$0.04  &  \\  
      88  &  4450  &	1.20	   &  1.50   &  --1.78$\pm$0.05   &   --1.74$\pm$0.05	&    --1.67$\pm$0.08  &  --1.46$\pm$0.04  &   \\ 
     221  &  4570  &	1.13	   &  1.40   &  --2.04$\pm$0.04   &   --2.04$\pm$0.04	&    --2.00$\pm$0.04  &  --1.86$\pm$0.04  &   \\ 
     224  &  4670  &	1.75	   &  1.40   &  --1.87$\pm$0.04   &   --1.78$\pm$0.04	&    --1.76$\pm$0.05  &  --1.47$\pm$0.04  &    \\	
  200005  &  3920  &	0.00	   &  2.20   &  --2.10$\pm$0.02   &   --1.92$\pm$0.06	&    --1.97$\pm$0.05  &  --1.74$\pm$0.05  &  \\  
  200006  &  3910  &	0.04	   &  2.10   &  --1.84$\pm$0.03   &   --1.78$\pm$0.05	&    --1.69$\pm$0.05  &  --1.55$\pm$0.04  &  \\  
  200025  &  4060  &	0.57	   &  1.90   &  --1.72$\pm$0.02   &   --1.75$\pm$0.05	&    --1.48$\pm$0.05  &  --1.45$\pm$0.04  &  \\ 
  200031  &  4290  &	0.72	   &  1.80   &  --1.96$\pm$0.03   &   --1.97$\pm$0.04	&    --1.85$\pm$0.05  &  --1.64$\pm$0.04  &  AGB? \\  
  200043  &  4300  &	0.73	   &  1.70   &  --1.94$\pm$0.04   &   --1.95$\pm$0.04	&    --1.86$\pm$0.04  &  --1.69$\pm$0.05  &  AGB? \\ 
  200068  &  4400  &	0.82	   &  1.60   &  --2.00$\pm$0.04   &   --2.00$\pm$0.04	&    --1.93$\pm$0.05  &  --1.74$\pm$0.04  &  \\  
  200076  &  4390  &	0.80	   &  1.60   &  --2.05$\pm$0.04   &   --2.06$\pm$0.04	&    --2.01$\pm$0.05  &  --1.82$\pm$0.04  &  \\  
  200080  &  4520  &	0.70	   &  1.70   &  --2.01$\pm$0.04   &   --2.00$\pm$0.03	&    --1.96$\pm$0.05  &  --1.78$\pm$0.04  &  AGB \\	
  200083  &  4440  &	1.20	   &  1.50   &  --1.73$\pm$0.04   &   --1.75$\pm$0.04	&    --1.60$\pm$0.05  &  --1.46$\pm$0.05  &  AGB\\  
  200101  &  4400  &	0.90	   &  1.50   &  --1.89$\pm$0.05   &   --1.91$\pm$0.05	&    --1.77$\pm$0.05  &  --1.64$\pm$0.05  &  \\  
  200104  &  4490  &	0.59	   &  1.70   &  --2.19$\pm$0.06   &   --2.19$\pm$0.03	&    --2.15$\pm$0.08  &  --2.00$\pm$0.03  &  AGB \\

\hline
          &        &	 	   &         &  --1.92$\pm$0.03   &   --1.91$\pm$0.03	&    --1.81$\pm$0.04  &  --1.64$\pm$0.04  &   \\	

\hline
\enddata 
\end{deluxetable}


\begin{deluxetable}{lcccccccc}
\tablecolumns{9} 
\tablewidth{0pc}  
\tablecaption{Atmospheric parameters, [FeI/H], [FeII/H], [TiI/H] and [TiII/H] abundances for the 
spectroscopic targets of M22, as derived with method (2). 
The last line lists the average abundances with the statistical error.}
\tablehead{ 
\colhead{Star} &   \teff\ & log~g & \vturb\ & [FeI/H] & [FeII/H] & [TiI/H] &  [TiII/H] & Notes\\
  &   (K)  &    &(km/s) & &  &  &  &}
\startdata 
\hline 
      51  &  4280  &	0.99	   &  1.70   &  --1.70$\pm$0.02   &   --1.72$\pm$0.04	&    --1.50$\pm$0.04  &  --1.41$\pm$0.05  &   \\ 
      61  &  4440  &	1.17	   &  1.60   &  --1.84$\pm$0.05   &   --1.72$\pm$0.04	&    --1.76$\pm$0.06  &  --1.49$\pm$0.05  &   \\ 
      71  &  4390  &	1.15	   &  1.60   &  --1.94$\pm$0.03   &   --1.78$\pm$0.04	&    --1.85$\pm$0.05  &  --1.51$\pm$0.05  &  \\  
      88  &  4470  &	1.30	   &  1.50   &  --1.77$\pm$0.05   &   --1.72$\pm$0.05	&    --1.66$\pm$0.08  &  --1.44$\pm$0.06  &   \\ 
     221  &  4640  &	1.81	   &  1.30   &  --2.00$\pm$0.05   &   --1.71$\pm$0.04	&    --1.98$\pm$0.06  &  --1.53$\pm$0.05  &   \\ 
     224  &  4650  &	1.80	   &  1.30   &  --1.88$\pm$0.05   &   --1.74$\pm$0.04	&    --1.79$\pm$0.07  &  --1.42$\pm$0.05  &    \\	
  200005  &  3900  &	0.30	   &  2.20   &  --2.07$\pm$0.03   &   --1.67$\pm$0.09	&    --2.07$\pm$0.10  &  --1.62$\pm$0.06  &  \\  
  200006  &  3960  &	0.34	   &  2.10   &  --1.80$\pm$0.04   &   --1.72$\pm$0.09	&    --1.63$\pm$0.10  &  --1.48$\pm$0.06  &  \\  
  200025  &  4070  &	0.63	   &  1.80   &  --1.70$\pm$0.04   &   --1.71$\pm$0.07	&    --1.45$\pm$0.08  &  --1.46$\pm$0.06  &  \\ 
  200031  &  4240  &	0.86	   &  1.80   &  --2.02$\pm$0.04   &   --1.81$\pm$0.05	&    --1.98$\pm$0.07  &  --1.53$\pm$0.05  &  AGB? \\  
  200043  &  4270  &	0.93	   &  1.70   &  --1.98$\pm$0.04   &   --1.77$\pm$0.05	&    --1.96$\pm$0.05  &  --1.56$\pm$0.06  &  AGB? \\ 
  200068  &  4340  &	1.16	   &  1.60   &  --2.10$\pm$0.05   &   --1.75$\pm$0.05	&    --2.11$\pm$0.07  &  --1.54$\pm$0.05  &  \\  
  200076  &  4410  &	1.24	   &  1.60   &  --2.05$\pm$0.04   &   --1.82$\pm$0.04	&    --2.03$\pm$0.05  &  --1.54$\pm$0.05  &  \\  
  200080  &  4570  &	1.28	   &  1.70   &  --2.03$\pm$0.05   &   --1.74$\pm$0.04	&    --1.99$\pm$0.06  &  --1.52$\pm$0.05  &  AGB \\	
  200083  &  4430  &	1.18	   &  1.60   &  --1.76$\pm$0.04   &   --1.76$\pm$0.05	&    --1.60$\pm$0.05  &  --1.48$\pm$0.05  &  AGB\\  
  200101  &  4480  &	1.37	   &  1.50   &  --1.83$\pm$0.05   &   --1.71$\pm$0.05	&    --1.71$\pm$0.06  &  --1.44$\pm$0.06  &  \\  
  200104  &  4520  &	1.36	   &  1.70   &  --2.26$\pm$0.05   &   --1.82$\pm$0.04	&    --2.23$\pm$0.07  &  --1.55$\pm$0.05  &  AGB \\	
\hline 
        &        &	 	   &         &  --1.92$\pm$0.04   &   --1.75$\pm$0.01	&    --1.84$\pm$0.05  &  --1.50$\pm$0.01  &   \\	

\hline
\enddata 
\end{deluxetable}


\begin{deluxetable}{lcccccccc}
\tablecolumns{9} 
\tablewidth{0pc}  
\tablecaption{ Atmospheric parameters, [FeI/H], [FeII/H], [TiI/H] and [TiII/H] abundances for the 
spectroscopic targets of M22, as derived with method (3). 
The last line lists the average abundances with the statistical error.}
\tablehead{ 
\colhead{Star} &   \teff\ & log~g & \vturb\ & [FeI/H] & [FeII/H] & [TiI/H] &  [TiII/H] & Notes\\
  &   (K)  &    &(km/s) & &  &  &  &}
\startdata 
\hline 
      51  &  4232  &	0.96	   &  1.70   &  --1.76$\pm$0.05   &   --1.72$\pm$0.06	&    --1.57$\pm$0.06  &  --1.37$\pm$0.06  &   \\ 
      61  &  4400  &	1.16	   &  1.60   &  --1.87$\pm$0.05   &   --1.73$\pm$0.05	&    --1.81$\pm$0.06  &  --1.50$\pm$0.07  &   \\ 
      71  &  4435  &	1.17	   &  1.50   &  --1.88$\pm$0.05   &   --1.82$\pm$0.05	&    --1.76$\pm$0.08  &  --1.50$\pm$0.08  &  \\  
      88  &  4537  &	1.32	   &  1.60   &  --1.71$\pm$0.06   &   --1.80$\pm$0.06	&    --1.53$\pm$0.06  &  --1.49$\pm$0.07  &   \\ 
     221  &  4737  &	1.84	   &  1.50   &  --1.90$\pm$0.05   &   --1.81$\pm$0.05	&    --1.84$\pm$0.07  &  --1.57$\pm$0.07  &   \\ 
     224  &  4746  &	1.84	   &  1.50   &  --1.80$\pm$0.04   &   --1.82$\pm$0.04	&    --1.65$\pm$0.07  &  --1.48$\pm$0.06  &    \\	
  200005  &  3992  &	0.34	   &  2.20   &  --2.08$\pm$0.04   &   --1.78$\pm$0.04	&    --1.88$\pm$0.06  &  --1.62$\pm$0.07  &  \\  
  200006  &  3986  &	0.36	   &  2.10   &  --1.80$\pm$0.05   &   --1.77$\pm$0.05	&    --1.56$\pm$0.06  &  --1.47$\pm$0.07  &  \\  
  200025  &  4116  &	0.65	   &  1.90   &  --1.68$\pm$0.05   &   --1.81$\pm$0.05	&    --1.36$\pm$0.08  &  --1.45$\pm$0.09  &  \\ 
  200031  &  4271  &	0.87	   &  1.80   &  --1.97$\pm$0.05   &   --1.88$\pm$0.05	&    --1.93$\pm$0.06  &  --1.54$\pm$0.07  &  AGB? \\  
  200043  &  4351  &	0.97	   &  1.70   &  --1.89$\pm$0.04   &   --1.84$\pm$0.05	&    --1.81$\pm$0.08  &  --1.58$\pm$0.08  &  AGB? \\ 
  200068  &  4409  &	1.19	   &  1.60   &  --2.00$\pm$0.05   &   --1.85$\pm$0.07	&    --1.98$\pm$0.06  &  --1.56$\pm$0.07  &  \\  
  200076  &  4475  &	1.27	   &  1.60   &  --1.98$\pm$0.05   &   --1.87$\pm$0.05	&    --1.92$\pm$0.06  &  --1.60$\pm$0.06  &  \\  
  200080  &  4618  &	1.30	   &  1.80   &  --1.95$\pm$0.05   &   --1.80$\pm$0.07	&    --1.90$\pm$0.06  &  --1.55$\pm$0.07  &  AGB \\	
  200083  &  4567  &	1.24	   &  1.60   &  --1.61$\pm$0.05   &   --1.84$\pm$0.05	&    --1.36$\pm$0.08  &  --1.49$\pm$0.06  &  AGB\\  
  200101  &  4527  &	1.39	   &  1.60   &  --1.79$\pm$0.06   &   --1.80$\pm$0.04	&    --1.62$\pm$0.06  &  --1.52$\pm$0.07  &  \\  
  200104  &  4655  &	1.41	   &  1.70   &  --2.04$\pm$0.07   &   --1.89$\pm$0.07	&    --1.99$\pm$0.07  &  --1.63$\pm$0.09  &  AGB \\	

\hline
          &        &	 	   &         &  --1.86$\pm$0.03   &   --1.81$\pm$0.01	&    --1.73$\pm$0.05  &  --1.52$\pm$0.02  &   \\	

\hline
\enddata 
\end{deluxetable}

\begin{deluxetable}{lcccc}
\tablecolumns{5} 
\tablewidth{0pc}  
\tablecaption{Abundance ratios for the s-process elements Y, Ba, La and Nd.}
\tablehead{ 
\colhead{Star}  & [YII/FeII] & [BaII/FeII] &  [LaII/FeII] & [NdII/FeII]\\
  }
\startdata 
\hline 
      51  &       +0.15$\pm$0.04   &   +0.65$\pm$0.07   &   +0.54$\pm$0.04  &	  +0.48$\pm$0.04     \\ 
      61  &      --0.37$\pm$0.05   &   +0.04$\pm$0.08   &   +0.08$\pm$0.04  &	  +0.08$\pm$0.04     \\ 
      71  &      --0.35$\pm$0.05   &   +0.14$\pm$0.08   &   +0.16$\pm$0.06  &	  +0.07$\pm$0.04    \\  
      88  &       +0.15$\pm$0.05   &   +0.66$\pm$0.08   &   +0.47$\pm$0.04  &	  +0.32$\pm$0.05     \\ 
     221  &      --0.42$\pm$0.04   &   +0.15$\pm$0.10   &   +0.11$\pm$0.04  &	  +0.00$\pm$0.05     \\ 
     224  &       +0.08$\pm$0.05   &   +0.51$\pm$0.07   &   +0.34$\pm$0.06  &	  +0.14$\pm$0.04      \\	
  200005  &      --0.42$\pm$0.06   &   +0.06$\pm$0.12   &   +0.03$\pm$0.05  &	 --0.10$\pm$0.05    \\  
  200006  &       +0.06$\pm$0.08   &   +0.42$\pm$0.11   &   +0.33$\pm$0.05  &	  +0.33$\pm$0.05    \\  
  200025  &       +0.09$\pm$0.05   &   +0.54$\pm$0.06   &   +0.43$\pm$0.06  &	  +0.39$\pm$0.05     \\ 
  200031  &      --0.35$\pm$0.04   &   +0.03$\pm$0.08   &   +0.03$\pm$0.06  &	 --0.01$\pm$0.04    \\  
  200043  &      --0.34$\pm$0.05   &   +0.04$\pm$0.07   &   +0.04$\pm$0.04  &	 --0.03$\pm$0.04     \\ 
  200068  &      --0.41$\pm$0.04   &   +0.04$\pm$0.08   &   +0.00$\pm$0.05  &	 --0.05$\pm$0.04    \\  
  200076  &      --0.37$\pm$0.04   &   +0.10$\pm$0.08   &   +0.19$\pm$0.05  &	 --0.02$\pm$0.04    \\  
  200080  &      --0.45$\pm$0.04   &   +0.13$\pm$0.09   &   +0.08$\pm$0.05  &	  +0.03$\pm$0.05     \\ 
  200083  &       +0.13$\pm$0.06   &   +0.72$\pm$0.08   &   +0.58$\pm$0.04  &	  +0.43$\pm$0.04    \\  
  200101  &       +0.12$\pm$0.06   &   +0.77$\pm$0.09   &   +0.53$\pm$0.04  &	  +0.37$\pm$0.05    \\  
  200104  &      --0.45$\pm$0.04   &   +0.07$\pm$0.09   &   +0.15$\pm$0.06  &	 --0.01$\pm$0.05     \\ 
\hline
 \hline
\enddata 
\end{deluxetable}

\end{document}